\documentclass[12pt,showpacs,showkeys,nofootinbib]{revtex4}
\usepackage{graphicx}
\usepackage{CJK}
\usepackage{psfrag}
\usepackage{amsmath}
\usepackage{amssymb}
\usepackage{bm}
\usepackage{bbm}
\begin{document}
\begin{CJK*}{GBK}{kai}
\title{Production of the tetraquark state $T_{cc}$ at the B-factories}
\author{Reyima Rashidin$^{1,2}$}

\address{$^1$
Key Laboratory of Frontiers in
Theoretical Physics, \\The Institute of Theoretical Physics, Chinese
Academy of Sciences, Beijing 100190, People's Republic of China\vspace{0.2cm}\\
$^2$ School of Physics Science and Technology, Xinjiang University,
Urumqi 830046, People's Republic of China\vspace{0.2cm}}



\begin{abstract}
{\bf Abstract}: We study  production  of the tetraquark state $T_{cc}$ via virtual photon at the
B-factories in the QCD factorization framework. We predict the cross section of tetraquark state
production in the leading order at the B-factories.
\end{abstract}

\pacs{13.66.Bc, 14.40.Rt, 12.39.jh, 13.87.fh}

\keywords{tetraquark, diquark, fragmentation, production}

\maketitle
\newpage

\section{Introduction\label{introduction}}
The exotic hadrons, such as tetraquark states, hybrid states, and open-charm meson etc, are always hot topic in hadronic physics. In
recent years, some new exotic hadrons, such as $X(3872)$, $Y_{b}(4260)$, and $Z^{\pm}(3895)$~\cite{ex001,ex002,ex003,ex004,ex005}, discovered in
the experimental. Their mass are closer to sum of two $D$ mesons~\cite{ex0005,ex006}. These hadrons are probably molecular
models in the tetraquark picture. In the molecular model of tetraquark states, two mesons compose and formes a tetraquark by attractive residual strong interaction. Except for molecular models,
tetraquark state has another configuration which is diquark-antidiquark model. In diquark-antidiquark model, the diquark formed
by two quarks is binding with antidiquark formed by two antiquarks.

Tetraquark states which are consists of the double heavy quarks and double light antiquarks as well
as their charge conjugates can be very good candidates. They can be regarded as diquark-antidiquark
model in the tetraqaurk picture. In these tetraquark states, the two heavy quarks could tightly
bounded and formed the color anti-triplet antisymmetric diquark cluster by strong attractive
interaction.\footnote{In the one-gluon exchange potential, when two quarks form to the color
anti-triplet, interaction between them are strong attractive interaction. When the two quarks
formed color sextet configuration, the interaction between  them are repulsive interaction. Since
they are mainly form to the color anti-triplet configuration.} The size of the diquark is about
 $ 0.1- 0.3  \textrm{fm}$ \cite{ccsize} and it can be considered as a point-like particle. It
contributes the color interactions to double light antiquarks as a heavy antiquark. The two light anti-quarks are bounded with heavy
diquark.  As a result, the two heavy quarks and two light antiquarks
can form a compact tetraquark state.

The spin of a tatraquark state is composition of the spins of the four quarks and the relative
orbital angular momenta between them. For $S$-waves, the all orbital angular momenta vanish. When
the two heavy quarks inside the diquark are identical, only spin one state for double heavy
diquark is allowed because of the flavor anti-symmetric character. While, when the two heavy quarks
in the double heavy diquark have different flavors, the spin of diquark may be one or zero. This
discussion also can be adopt to the two light antiquarks sector. In this paper, we just consider
the tetraquark states which are consists of two charm quarks and two light anti-quarks, with
all orbital angular momenta vanished.  We denote this tetraquark states by $T_{cc}$.

  Flavor features of tetraquark with two heavy quarks and two light anti-quarks are very different
from the conventional hadrons. If they are founded in experiments, we undoubtedly confirm the existence of the tetraquark states. The production of these tetraquark states have been studied in theoretically with different ways~\cite{th001,th002,th003,th004,th005,th006,th007,th008}.
 In paper~\cite{th001}, authors studied the production of double heavy tetraquark states at the Large Hadron Collider(LHC) and predicted
possible decay channels of these tetraquark states. Their results showed that there are very good opportunity to observe double heavy
tetraquark states at the LHC. One of the motivation of this work is to investigate if there is any possibility to observe double heavy
tetraquark states at the B-factories. The other one is that the hadronic background at the $e^{+}e^{-}$ colliders is simpler than
 at the  proton-proton collider.

In this paper, we calculate the total cross section of tetraquark states $T_{cc}$ via virtual
photon at the B-factories. We present the cross section at the leading order in $\alpha_{s}$ in
$e^{+}e^{-}$ annihilation. The process can be factorized into three subprocess. The first
subprocess is that the $e^{+}e^{-}\to c+c+\bar{c}+\bar{c}$, which is a hard process at the distance
scale $1/m_{Q}$ and it can be calculated by perturbative QCD(pQCD). For the second subprocess,
formation of qiduark from two heavy quarks, the relative momentum between two heavy quarks is small
with $m_{Q}v\ll m_{Q}$. Such subprocess should be calculated by non-perturbation methods. At the subprocess, the tetraquark will be
produced via the diquark picking up two light antiquarks from the vacuum at distance scale $~1/\Lambda_{QCD}$
 and it is also a non-perturbative process which can be described by fragmentation model.

\section{the calculation for production of $T_{cc}$}
\label{production}
\begin{figure}[htbp]
\begin{center}
\includegraphics*[scale=0.7]{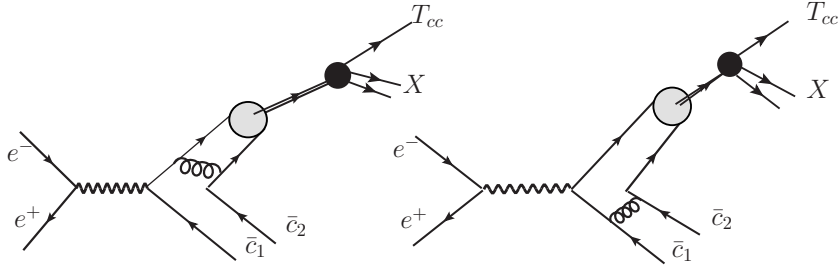}

\caption{Typical Feynman diagrams for the process $e^{+}e^{-}\to
T_{cc}+X$.}
\label{cfour}
\end{center}
\end{figure}

 In the B-factory, the tetraquark particles production via a virtual photon $\gamma^{*}$  as shown
in Fig.\ref{cfour}. As discussed above, in the production process of the tetraquark states
$T_{cc}$, these distance scales satisfy the hierarchy relation
$1/m_{Q} \ll 1/(m_{Q}v) \ll 1/(m_{Q}v^2)$. Different physical processes happening at those
distinct energy scales are factorized. They can be calculated by  pQCD, NRQCD, and the
fragmentation function respectively. Therefore, total cross section can described as


\begin{eqnarray}
\sigma(e^{+}e^{-}\to T_{cc}+X)&=&\frac{1}{2S}\frac{1}{4}\frac{1}{d!}\int_{0}^{1}dx N\delta(1-x)
\nonumber\\
&.&|\frac{\Psi_{cc}(0)}{\sqrt{d!}}|^{2}\int d\Pi_{3}|C(\alpha_{s},cc)|^{2},
\label{xsec}
\end{eqnarray}

where$\frac{1}{4}$ is the average spin of initial particle; $d!$ is from permutation symmetry
between identical final particles; $S$ is center of mass energy of initial electron and positron
and the masses of electron and positron are ignored;  $C(\alpha_{s},cc)$ is the short distance
coefficient describing the production rate of the color anti-triplet point-like $cc$ state at
the energy scale $m$ or higher; $\Psi_{cc}(0)$ is the wave function at the origin of the $S$-wave
diquark state; $N\delta(1-x)$ is the fragmentation function of the diquark into color-singlet
tetraquark state $T_{cc}$ and $N$ is the normalization factor in the fragmentation function;
$d\Pi_{3}$ is the Lorentz invariant three-body phase space integral element, which is
\begin{eqnarray}
d\Pi_{3}=(2\pi)^{4}\delta^{4}(p_{e1}+p_{e2}-P-q_{1}-q_{2})\frac{d^{3}P}{(2\pi)^{3}2E}
\frac{d^{3}q_{1}}{(2\pi)^{3}2E_{1}}\frac{d^{3}q_{2}}{(2\pi)^{3}2E_{2}}.
\end{eqnarray}


The short distance coefficient is calculated by perturbative QCD and
can be expanded as a powers of $\alpha_{s}$ at energy scale m or
higher. For the Formation of diquark from two heavy quarks, where
the motion between two heavy quarks is very slow at distance
$1/m_{Q}v$, it is calculated by the potential model. The tetraquark
state will be produced by diquark picking up two antiquarks from the
vacuum at distance scale $1/\Lambda_{QCD}$. It is also a
non-perturbative process which can be calculated by fragmentation
function.

\subsection{The calculation of short distance coefficient $C(\alpha_{s},cc)$}

The short distance coefficient $|C(\alpha_{s},cc)|^{2}$ is
proportional to squared matrix element $M'(e^{+}e^{-}\to
cc+\bar{c}+\bar{c})$ at the tree level, where $cc$ is design to be
color anti-triplet and spin one point-like state, which is composed
by two charm quarks. The dependence of scattering amplitude including
Clebsh-Gorden coefficients are written in terms of fermion string as
 Ref.\cite{pro001}.


 For convenience of calculation, the wave function of the diquark was described as projection
 operator in non-relativistic approximation. Correspondingly, we reorganized the amplitude of
 subprocess $e^{+}e^{-}\to cc(P)+\bar{c}(q_{1})+\bar{c}(q_{2})$. There are eight Feynman diagrams with
 four different topological structures, since we just need to calculate four amplitudes. They are

\begin{eqnarray}
M_1^{\mu}&=& \frac{1}{(\frac{P}{2}+q_1)^2}\bar{u}(q_1,s_3)\gamma^{\nu}  \Pi_{3}(P,\lambda) \gamma_{\nu}\frac{( /\!\!\!P+/\!\!\!q_1+m_c)}{(P+q_1)^2-m_{c}^2}\gamma^{\mu}  v(q_2,s_4), \nonumber\\
M_2^{\mu}&=&-\frac{1}{(\frac{P}{2}+q_2)^2}\bar{u}(q_1,s_3)\gamma^{\mu} \frac{(/\!\!\!P+/\!\!\!q_2+m_c)}{(P+q_2)^2-m_{c}^2}\gamma^{\nu}\Pi_{3}(P,\lambda)\gamma_{\nu}v(q_2,s_4),\nonumber\\
M_3^{\mu}&=&\frac{1}{(\frac{P}{2}+q_1)^2} \bar{u}(q_1,s_3)\gamma^{\nu}  \Pi_{3}(P,\lambda)\gamma^{\mu}\frac{(\frac{ /\!\!\!P}{2}+/\!\!\!q_1+/\!\!\!q_2-m_{c})}{(\frac{P}{2}+q_1+q_2)^2-m_{c}^2}\gamma_{\nu} v(q_2,s_4), \nonumber\\
M_4^{\mu}&=&-\frac{1}{(\frac{P}{2}+q_2)^2}\bar{u}(q_1,s_3) \gamma^{\nu}\frac{(\frac{/\!\!\!P}{2}+/\!\!\!q_1+/\!\!\!q_2-m_{c})}{(\frac{P}{2}+q_1+q_2)^2-m_{c}^2}\gamma_{\mu}\Pi_{3}(P,\lambda)\gamma_{\nu} v(q_2,s_4).
\label{4am}
\end{eqnarray}
then
\begin{equation}
M'=\sum_{a=1}^{4}ig_{s}^{2}g_{e}^{2}e e_{Q'}c_{1} L_{\mu}M_{a}^{\mu},
\end{equation}
where $e_{Q'}$ and $e$ are electric charge of charm quarks and
initial particles in the unit $e$;
$c_{1}=\frac{\varepsilon^{ijk}}{4\sqrt{2}}\lambda^{c}_{ij}\lambda^{c}_{mk}$
is color factor,
 $\varepsilon^{ijk}$ is the anti-symmetry tensor ; $g_{s}$ and $g_{e}$ are strong coupling constant
 and electromagnetic constant; $L_{\mu}=\frac{1}{S}\bar{v}(p_{e2})\gamma_{\mu} u(p_{e1})$ is initial
  leptonic part which includes photon propagator factor; $m_{c}$ is mass of charm quark ; $P$  is
  four momentum of point-like state $cc$; $q_{1}$, $q_{2}$ and $s_{3}$, $s_{4}$ are four momentum
  and spin of anti charm quarks ; $\Pi_{3}$ is projection operator of color anti triplet and spin
  one point-like state of two charm quarks. In principle, the projection wave function for binding
  of $cc$  should written as

\begin{equation}
\Pi_{3}=\sum_{s_{1},s_{2}}<\frac{1}{2},s_{1z};\frac{1}{2},s_{2z}\mid S,S_{z}> \bar{u}(p_1,s_1)
\bar{u}(p_2,s_2).
\label{3pp}
\end{equation}

where $\bar{u}(p_{1}, s_{1})$ and $\bar{u}(p_{2},s_{2})$ are spinor of  two charm quarks inside the
point-like state $cc$; $p_{1}$, $p_{2}$ and $s_{1}$, $s_{2}$ are four momentum and spin of these
 charm quarks, respectively; $<\frac{1}{2},s_{1z}; \frac{1}{2},s_{2z}\mid S,S_{z}> $ is Clebsh-Gorden
 coefficient of the coupling of two charm quarks $cc$.

If we represent one of the quarks by the charge conjugate field, the constructions look more
familiar : $qq\to \bar{q}_{c}q$, where $\bar{q}_{c}= -iq^{T}\gamma^{2}\gamma^{0}$. In the paper \cite{wuxg}, the authors transform the fermion line
$(\bar{u}(p_{1},s_{1})\Gamma v (q_{1},s_{3}) $ of the diquark - production case to anti-fermion
line of the meson-production $(v(p_{1},s_{1})\Gamma' \bar{u}(q_{1},s_{3}))$ by using characters of
charge conjugation operator.\footnote{ To obtain the projection operator of diquark, we used some
characters of the charge conjugation operator $\rm{C}=-i\gamma^{2}\gamma^{0}$, i.e $CC^{-1}=1$,
$v(p,s)=C\bar{u}^{T}(p,s)$, and $C\gamma^{\mu}C^{-1}=-(\gamma^{\mu})^{T}$. See Ref.\cite{wuxg}for
 further discussion.} Similarly

\begin{equation}
\Pi_{3}=\sum_{s_{1},s_{2}}\bar{u}(p_2,s_2)v({p_{1},s_{1}})<\frac{1}{2},s_{1z};\frac{1}{2},s_{2z}
\mid S,S_{z}>.
\end{equation}

Given four momentum of the diquark $P$, the relative four momentum
between two charm quarks $q$, the four momentum of two charm quarks
$p_{1}$, $p_{2}$ can be expressed as 

$$ p_1=\frac{m_c}{M_{cc}}P+q,~ p_{2}=\frac{m_c}{M_{cc}}P-q,$$
here $M_{cc}$ is mass of diquark. In the heavy quark limit, $M_{cc}=2m_{c}$, $P$ and $q$ are
orthogonal: $P.q=0$. In the rest frame of the diquark,

$$P=(2E,0),~q=(0,\vec{q}),~p_{1}=(E, \vec{q}), ~p_{2}=(E, -\vec{q}),$$

In the rest frame of the diquark, the amplitude is expanded in powers of relative velocity $v_c$
between two charm quarks. In this paper, we only consider the leading order contribution of $v_{c}$,
since the dependence of short distance coefficients to the relative momentum $q$ can be neglected.
 Therefore, the four momentum of charm quarks can be $p_1=p_2=\frac{1}{2}P$.

 The projection operator for the coupling of two charm quarks, under the non-relativistic
 approximation, takes the simple form~\cite{petreill}:
\begin{subequations}
\label{spin-pro-2}
\begin{eqnarray}
\Pi_1(P)&=&\frac{1}{2\sqrt{2m_c}}
\gamma_5 (/\!\!\!P+2m_c),\\
\Pi_3(P,\lambda)&=&\frac{1}{2\sqrt{2m_c}}
/\!\!\!\epsilon(\lambda) (/\!\!\! P +2m_c),
\end{eqnarray}
\end{subequations}

where $\epsilon^{\mu}(\lambda)$ is polarization vector for spin-triplet state. $\Pi_{1}$ and
$\Pi_{3}$ are projection operator for spin 0 and spin 1 state respectively.

\subsection{Formation of diquark from double heavy quarks}

According to the discussion in Ref.\cite{th001}, two charm quarks bind to form a stable
color-anti-triplet diquark. In the heavy quark limit, the distance between two charm quarks is
much smaller than $\frac{1}{\Lambda_{QCD}}$, since the $\psi_{cc}(0)$ is determined by Coulomb
potential~\cite{colomb001}. It can be predicted by solving shr\"{o}dinger equation with Coulomb
potential, $v(r)=-2\alpha_s/(3r)$. For anti-triplet $cc$ diquark state, its numerical value is
$|\psi_{cc}(0)|^2=0.0198\textrm{GeV}^3$. And the production amplitude of diquark via $e^{+}e^{-}$
annihilation expressed as

\begin{eqnarray}
A&=& \sum_{s_{3},s_{4}}\sum_{color}\frac{\sqrt{2M_{cc}}}{\sqrt{2m_{c}}\sqrt{2m_{c}}}\frac{\psi(0)}
{\sqrt{d!}}(-iM')\nonumber\\
&=&C(\alpha_{s},cc)\frac{\psi(0)}{\sqrt{d!}},
\label{3pp}
\end{eqnarray}

where $C(\alpha_{s},cc)$ includes  tree level amplitude of subprocess
$e^{+}e^{-}\to cc+\bar{c}+\bar{c}$ and  normalization factor in diquark wave function.

\subsection{diquark fragmentation into tetraquark state}

As discussed above, in the heavy quark limit, the heavy diquark
cluster takes a  point-like anti-quark color structure. In
principle, at the low energy scale, quark can omit arbitrary number
soft-gluons and these gluons can bound heavy diquark with another
two light anti-quarks which picked up from the vacuum. We can
simulate this subprocess $ cc_{\bar{3}} \to T_{cc}$ by fragmentation
mode of the heavy quark. Because the symmetry between the heavy
quarks obtain two light quarks from vacuum by fragmentation process
to antiquark fragmentation, we use the fragmentation mode of diquark
with the one of the two heavy quarks.

The fragmentation model to produce the tetraquark are nonperturbative(scale is
$\frac{1}{\Lambda_{QCD}}$). In this paper, we choose one of the most commonly used fragmentation
function, the $\delta$ function model~\cite{peterson}. The fragmentation function is


\begin{eqnarray}
\label{saqu}
D(x)=N\delta(1-x),
\end{eqnarray}

Where $N$ is the normalization constant. The normalization constant can be fitted using the
constraint condition.
\begin{eqnarray}
\label{saqu}
\int dx {D_{(cc)}(x)}=R(cc_{\bar{3}}\to T_{cc}).
\end{eqnarray}
  The fragmentation probability of charm quark fragmentation into baryon is
$R[c\to \Lambda_{c}]=0.094\pm 0.035$ which  is measured by the
$e^{+}e^{-}$ colliders ~\cite{fragment}. Thus, we approximately take
the fragmentation probablity of double heavy diquark fragmentation
into tetraquark with double charm quarks and two light antiquarks as
$R(cc_{\bar{3}}\to T_{cc})= 0.1$.

\section{result and summary}
\label{summary}
In this work, we studied production of the tetraquark states at the B-factories. And we predicted
the production cross section of the process $e^{+}e^{-}\to T_{cc}+X$  at the leading order of
$\alpha_{s}$ expansion at the KEKB.  In order to obtain numerical result, we choose the parameters
of the fine structure constant and strong coupling constant as $\alpha=\frac{1}{131}$,
$\alpha_{s}(2m_{c})=0.24$, and mass of charm quarks and double charm diquark are
$m_c=1.5 \textrm{GeV}$, $M_{cc}=m_{c}+m_{c}=3.0 \textrm{GeV}$, and the center mass energy  is
$\sqrt{S}=10.58 \textrm{GeV}$ at the KEKB.

$$\sigma_{total}= 2.669 \pm 0.0295 \textrm{fb}$$

 Result shows that it is difficult to search tetraquark states at the normal $e^{+}e^{-}$ colliders.
But there may be some hope to observe the tetraquark states at the
high facility B-factories, like super-B at the KEK-B, it's
luminosity will be promoted to $8\times
10^{35}\textrm{cm}^{-2}.\textrm{s}^{-1}$ , and it implies that
around $65000$ events can be produced at this collider one year.
Although there may be difficulties for the discovery of these
tetraquark states in this collider, once the detection efficiency is
considered.


\begin{acknowledgments}
R. R. was thanks Prof. Yu-Qi Chen and Dr. Su-Zhi Wu for helpful
discussion. This work was supported by the National Natural Science
Foundation of China under grants No.~11275242 and No.~11165014.

\end{acknowledgments}

\end{CJK*}
\end{document}